\documentclass{sig-alternate-10pt}
\usepackage{graphicx}
\usepackage{algorithm}
\usepackage{algorithmic}

\begin{document}
\title{Spiral Walk on Triangular Meshes : Adaptive Replication in Data P2P Networks.}

\numberofauthors{3} 
\author{
\alignauthor Nicolas Bonnel\\
       \affaddr{VALORIA}\\
       \affaddr{European University of Brittany}\\
       \affaddr{Vannes, France}\\
       \email{nicolas.bonnel@univ-ubs.fr}
\alignauthor Gildas Menier\\
       \affaddr{VALORIA}\\
       \affaddr{European University of Brittany}\\
       \affaddr{Vannes, France}\\
       \email{gildas.menier@univ-ubs.fr}
\alignauthor Pierre-Francois Marteau\\
       \affaddr{VALORIA}\\
       \affaddr{European University of Brittany}\\
       \affaddr{Vannes, France}\\
       \email{pierre-francois.marteau@univ-ubs.fr}
}

\maketitle

\begin{abstract}
We introduce a decentralized replication strategy for peer-to-peer file exchange based on exhaustive exploration of the neighborhood of any node in the network. The replication scheme lets the replicas evenly populate the network mesh, while regulating the total number of replicas at the same time. This is achieved by self adaptation to entering or leaving of nodes. Exhaustive exploration is achieved by a \textit{spiral walk} algorithm that generates a number of messages linearly proportional to the number of visited nodes. It requires a dedicated topology (a triangular mesh on a closed surface). We introduce protocols for node connection and departure that maintain the triangular mesh at low computational and bandwidth cost. Search efficiency is increased using a mechanism based on dynamically allocated super peers. We conclude with a discussion on experimental validation results.
\end{abstract}




\section{Introduction}

In the last few years, the quality (reliability and bandwidth) of internet connections has increased drastically. With DSL, users have today an almost permanent high bandwidth access to the net. During the download of data, the transfert rate is now only limited by the capability of the server to manage a large amount of high bandwith clients at the same time. Thanks to the peer-to-peer (P2P) mechanisms, each user can contribute to this data exchange mechanism as a potential server: this extends the sharing capability and features some fault resilient properties because of the data replication and distribution involved.

Furthermore, P2P architectures allow users to share resources. Few existing architectures allow CPU cycle sharing \cite{distnet}, but most of them are designed to share memory and to behave as distributed databases. Data replication keeps information available when a node fails or leaves the network. Replication is useful for scalability because if a lot of users want to have simultaneous access to the same information, a single peer may not be able to supply all the queries.

Several architectures have been proposed for P2P networks. Napster \cite{Napster}, for instance, is a centralized P2P network that was very popular in the early 00's. However the use of a central repository to answer queries makes this system poorly scalable and vulnerable to failure. Many others P2P systems do not rely on a central server: they are decentralized, thus they are very scalable, and fault tolerant.

Structured P2P networks associate network topology and location of data. Most of them implement a Distributed Hashtable (DHT) \cite{chord2001,ratnasamy00scalable,rowstron01pastry,zhao01tapestry,maymounkov02kademlia} and provide one basic operation: mapping a key to a specific node. This is performed by using a distributed hash function. Content routing is used to forward the key to the corresponding node. This strategy is very well suited for rare information retrieval (i.e. with a low number of replicas). Most of the structured P2P networks can retrieve a key in $O(log(N))$ messages in a network containing $N$ nodes. Their main limitation relies in poor performances for ranged and approximative queries, because hashing destroys the order on keys. Oceanstore \cite{oceanstore-asplos}, Cooperative File System (CFS) \cite{cfs:sosp01} or Glacier \cite{haeberlen-2005-glacier} are examples of storage systems, built on top of DHTs and on replicated data to provide high availability.

Conversely, unstructured P2P networks do not feature any constraint between the location of data and the network topology. Gnutella \cite{gnutella}, eDonkey 2000 \cite{eDonkey2000}, KaZaA \cite{Kazaa} and Freenet \cite{clarke01freenet} are examples of such working architectures. The previous versions of Gnutella use flooding: this generates a number of messages that increases exponentially with the radius of the search. The trafic generated by this mechanism limits the scalability. To solve this problem, \textit{random walk}, and variants such as \textit{k-random walk} \cite{lv01search}, have been proposed. If a node receives a query and cannot answer it, the query is forwarded to a randomly chosen neighbor. When a query is answered, it is then back-propagated to the node that initiated the query using the inverse path. Because random walk does not rely on the content of the query, much more complex queries can be processed, such as approximative or range queries.

A Gnutella-like system has been proposed in \cite{chawathe03making}. The system relies on measures performed by \cite{saroiu03measuring} that show the heterogenity in capabilities of the peers involved in an unstructured P2P network such as Gnutella. This system uses a biased random walk that visits the nodes with a higher degree first. Each node has pointers on data hosted by its neighbors, and each new node that enters the network tries to connect to a high-degree node. These features increase the overall performance of the system by three to five orders of magnitude.

In Freenet, routing efficiency increases as the network becomes more and more organized. However, it uses content routing of keys, which means that a user looking for particular data needs the knowledge of the key of the data. As for structured P2P systems, it is very costly to perform range or approximative queries.

In popular P2P systems like Gnutella, eDonkey 2000 or KaZaA, replication is performed by users that download copies of files on their own computer. This owner replication scheme is well suited for popular data, however rare data remains very hard to reach.

This paper presents the design of a P2P file sharing system called Adapnet that offers high availability in exchange for storage space offered by each user. Because it does not rely on hashing like DHT's or Freenet, our system can handle very complex queries including ranged or approximative queries. A loose structure allows adaptive replicating algorithms to execute while a super-peer architecture \cite{Yang.ICDE.2003} speeds up query answering and makes the system scalable.

Section 2 briefly describes our system. Section 3 presents the adaptive replication algorithm built on a dedicated topology and a well-suited marching algorithm called \textit{spiral walk}. Section 4 describes the architecture used for fast query answering. In the last section, we evaluate Adapnet on a simulated network designed on real P2P statistical features \cite{saroiu03measuring,1217970}.


\section{Overview}
In this article, we present a data P2P architecture designed to offer high availability. In this framework, users offer some disk space used by the system to store information replicas. The proposed system is designed to manage rare data: therefore mostly replicas of rare data are created and distributed across the network. Replicating rare data keeps a fixed amount of information reachable even if a significant amount of nodes are offline.

We introduce an algorithm designed to keep the distribution of the replicas as evenly spaced as possible to ensure that they remain in the smallest neighborhood possible of any node. The replication scheme involves an autonomous mechanism that adapts the number of replicas and their distribution in the network to the number of active nodes. For each piece of data, the quantity of replicas is lineary proportional to the overall number of nodes in the network. This quantity in the system is globaly unknown.

The replication scheme we propose relies on the exhaustive exploration of the neighborhood of any node. While some structured P2P architectures allow an easy exploration of this neighborhood, they only support queries on key: range or approximative queries are very costly or not supported at all. These topologies are delicate to maintain when nodes leave the network without notification.

Unstructured P2P architectures allow wide query expressivity but are slower that structured P2P networks to locate information. Using the heterogeneity among peer capabilities in the network, super peer networks can be built to speed up information retrieval. Because there is no heavy topological properties to maintain, these networks are very well suited for highly dynamical environments. However, this lack of topological properties makes the exhaustive exploration only possible with flooding, therefore very costly.

We present in this article a marching algorithm called \textit{spiral walk} that explores the neighborhood of any node generating a number of messages lineary proportional to the number of visited nodes. This algorithm needs a dedicated topology to perform. More precisely the topology required is a triangular mesh on a closed surface.

We describe simple protocols for node connection or topology repairment after a node failure or a node departure that maintain this loosely structured topology. Nodes connection or failure are handled by the network with constant processing time operations according to the overall number of nodes. The average additionnal connectivity introduced is low : nodes have an average degree of six.

Our system uses the heterogeneity featured in unstructured P2P networks described in \cite{saroiu03measuring} to speed up information retrieval. Peers with higher ressource (bandwidth, CPU, memory, ...) additionnaly compose a super-peer network used to efficiently forward queries. Its random topology makes this super-peer network have a low diameter. Super-peers hold indexing information of other nodes, allowing fast query answering.


\section{Adaptive Replication}
\subsection{Replication scheme}
\subsubsection{Overview\label{replicat}}
We introduce a replication scheme that adapts the number of data replica without knowing the overall number of active nodes in the network.

\begin{itemize}
 \item Let $\delta(n_1,n_2)$ be the distance between two nodes $n_1$ and $n_2$ in the network. $\delta(n_1,n_2)$ is the number of hops of the shortest path that links $n_1$ to $n_2$.
 \item Let $R_{n,r}=\{n_i|\delta(n_i,n)=r\}$.
 \item Let $d_i$ be a data and $D_n$ be the set of data hosted on a node $n$.
\end{itemize}


Efficient storage management requires the optimization of the number of replicas. We propose a self organizing framework in which the replicas move across the network, repulsing each other until a near to optimal placement has been found to fulfill the bounded distance requirement.

\begin{figure}[htbp]
\begin{center}
\includegraphics[scale=0.5]{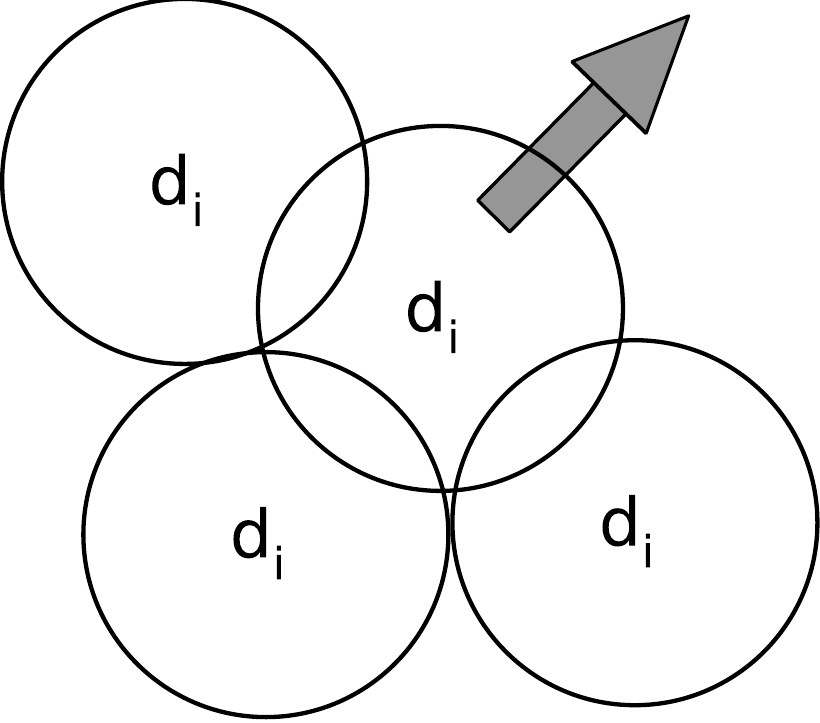}
\caption{$d_i$ replicas repulse each other to populate evenly the network.}
\end{center}
\end{figure}

This scheme involves competition between replicas and a cloning mechanism, which lets replicas homogeneously populate any region of the network. The main idea is very similar to the algorithms used in the artificial life or automata processing \cite{Langton1986}. The set of replicas can been seen as a population sharing a vital space by means of reproductive rewards and death penalties. This strategy is decentralized allowing the replicas to act as autonomous agents. This population reacts at runtime and adapts its distribution accross the network to the changes of environnement such as topological variations (entering or leaving of nodes) or resources variations (change of local storage space). Since the distribution involves a metric to compute a distance, we propose a dedicated topology, as well as a spiral walk strategy designed to fully and efficiently scan all the nodes below a maximal distance of any given node $n$.

\subsubsection{Replication scheme\label{replischeme}}
We present a replication scheme that is fully distributed and that maintains a number of replicas proportional to the size of the network, without knowing its size. The system mostly creates replicas of rare data. Replicas are spread accross the network, and periodically, a score related to the local density of replicas for a given $d_i$ is computed. According to this score, the replica are destroyed, moved, duplicated or kept on the node, as described in algorithm \ref{replication}. Because the considered topology is closed, replicas can repulse each other and become evenly spread accross the network.

\begin{algorithm}[htbp]
\caption{Replication from node $n$\label{replication}}
\begin{algorithmic}
\FORALL{$d \in D_n$}
\STATE $S \leftarrow Sc(n,d,r,\emptyset)$
\IF{$S>maxScore$}
\STATE Remove $d$ from $n$.
\ELSIF{$S=0$}
\STATE Let $\overline{R_{n,r,t}}$ a random subset of $R_{n,r}$ of size $t$.
\STATE Replicate $d$ on the node $n_p \in \overline{R_{n,r,t}}$ which minimizes $Sc(n_p,d,r,\emptyset)$. 
\ELSE
\STATE Let $n_{min} | \forall n_i \in R_{n,1},$
\STATE $Sc(n_{min},d,r,\{n\})\leq Sc(n_i,d,r,\{n\})$.
\STATE Move $d$ to $n_{min}$.
\ENDIF
\ENDFOR
\end{algorithmic}
\end{algorithm}

\begin{itemize}
 \item $I(n_i,d)=1$ if $d\in D_{n_i}$, $0$ otherwise
 \item $Sc(n,d,r,\Omega)=\sum_{n_i\in \chi}{I(n_i,d).(r-\delta(n,n_i)+1)^2}$
 \item $\Omega$ is the set of nodes to ignore for score computation 
 \item $\chi=\{n_i|\delta(n,n_i)\leq r\} - \Omega$
\end{itemize}

The choice of $maxScore$ is a trade-off between reactivity to the changes in the network and stability. A low value for $maxScore$ makes the number of replicas oscillate when a small variation of the number of nodes occurs. A higher value for $maxScore$ reduces the oscillations but lowers the reactivity of the system to adapt to the variation of its environment.

$t$ is an exploring constant. The higher $t$, the better the exploration for the placement of new replicas, however the associated computing cost also increases.

$r$ is the repulsive radius. The higher $r$, the lower the density of replicas.

Carrying the whole data during the scan of the neighborhood of a node can be costly. Therefore only hash values of those data are carried and compared. Moreover, many data replicas can be scanned simultaneously. Scanning only a part of the neighborhood leads to an approximative score value and produces oscillation of the quantity of replicas, even in a stable environment. This increases the convergence time of the replication algorithm. To overcome this limitation, we propose an efficient and exhaustive exploration of the neighborhood using spiral walk, described in section \ref{sw}.

When a replica has to be moved or replicated, the algorithm \ref{replication} makes the assumption that there will be enough space on the node to perform the replication. However, this cannot be garanteed in practice and there is a need to delete replicas when the cache of a node is full. This is performed by computing the score for all replicas and deleting the ones having the highest scores. If the score is the same for all replica (for exemple null), then replicas choosen at random are deleted from the cache.

\subsection{Support topology \label{topology}}
Replicas repulse each other on a surface. A ring topology could be maintained at low cost, however replicas would diffuse lineary. On the other hand, performing this repulsion on a higher dimensional space would speed up this diffusion, but it would be more costly to maintain the topology. The network topology we porpose is a triangular mesh on a closed 2-manifold offering a good trade-off between speed diffusion and maintening cost.

Using the Gauss-Bonnet theorem \cite{Stillwell1992} and the Euler characteristic, we deduce that the local curvature of the network topology around a node $n$ is:
\begin{itemize}
 \item positive if $n$ has five neighbors or less.
 \item null if $n$ has exactly six neighbors.
 \item negative if $n$ has seven neighbors or more.
\end{itemize}

As more and more nodes connect to the network, the average node degree tends to six, no matter the value of the initial average degree. This reduces the overall curvature of the topology. However, connection and failure recovery protocols locally deform the topology. Unfortunately, the presence of deformations in the topology alters our replication algorithm, as it reduces the possibilities for organizing replicas, and leads to a less efficient placement of replicas compared to a nearly flat topology. Replica placement is a key factor for the performances of the system, therefore we propose an algorithm to flatten the topology.

We assume that each node on the network has a unique identifier, this could be for instance the \textit{ip} address. We assume there is an order relationship between all identifiers. Nodes refresh their knowledge of their neighborhood by sending the identifier of their neighbors to their other neighbors. This can be viewed as a more elaborated \textit{ping} message. This knowledge stored on nodes speeds up common neighbors discovery for a pair of neighbor nodes.

\subsubsection{Topological invariant\label{invariant}}
In the triangular mesh, each edge belongs to exactly two triangles. More precisely :
\begin{center}
$\delta(n_1,n_2)=1 \Longrightarrow |R_{n_1,1}\cap R_{n_2,1}|=2$
\end{center}
In other words, if $n_1$ and $n_2$ are neighbors, then they have exactly two neighbors in common. This leads to a network in which any node has a minimal degree of four. The only network that fits those requirements with the lowest minimal degree is a tetrahedron.

\subsubsection{Node connection}
\begin{figure}[htbp]
\begin{center}
\includegraphics[scale=0.4]{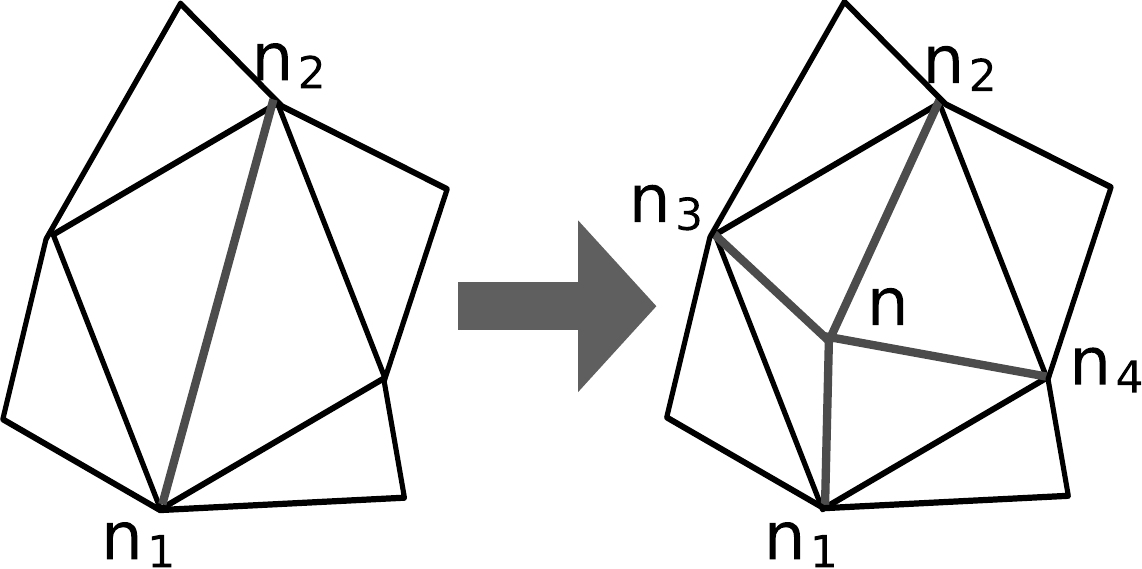}
\caption{\label{connex}Connexion of a new node to the network.}
\end{center}
\end{figure}

A node $n$ joining the network contacts a random node $n_1$ in the network. The way this random node $n_1$ is obtained can be achieved in various ways (for instance with a central server registering few nodes taken at random) and is not addressed here. A node $n_2$, neighbor of $n_1$ is then taken at random. Connection between $n_1$ and $n_2$ is terminated and new connections are created between $n$, $n_1$ and $n$, $n_2$. Finally, $n$ contacts the two neighbors common to $n_1$ and $n_2$, according to the property introduced in section \ref{invariant}, and creates a connection with them, as described in figure \ref{connex}.

This connection protocol adds one node and three connections to the network, bringing the average node degree in the network closer to six. This increases local curvatures in the topology, as $n$ has a degree of four and $n_3$ and $n_4$ have their degree increased by one. That is why the node $n$ then tries to acquire new neighbors by flattening the topology locally to its neighborhood. This flattening process is described in section \ref{flattening}.

\subsubsection{Node disconnection}
\begin{figure}[htbp]
\begin{center}
\includegraphics[scale=0.3]{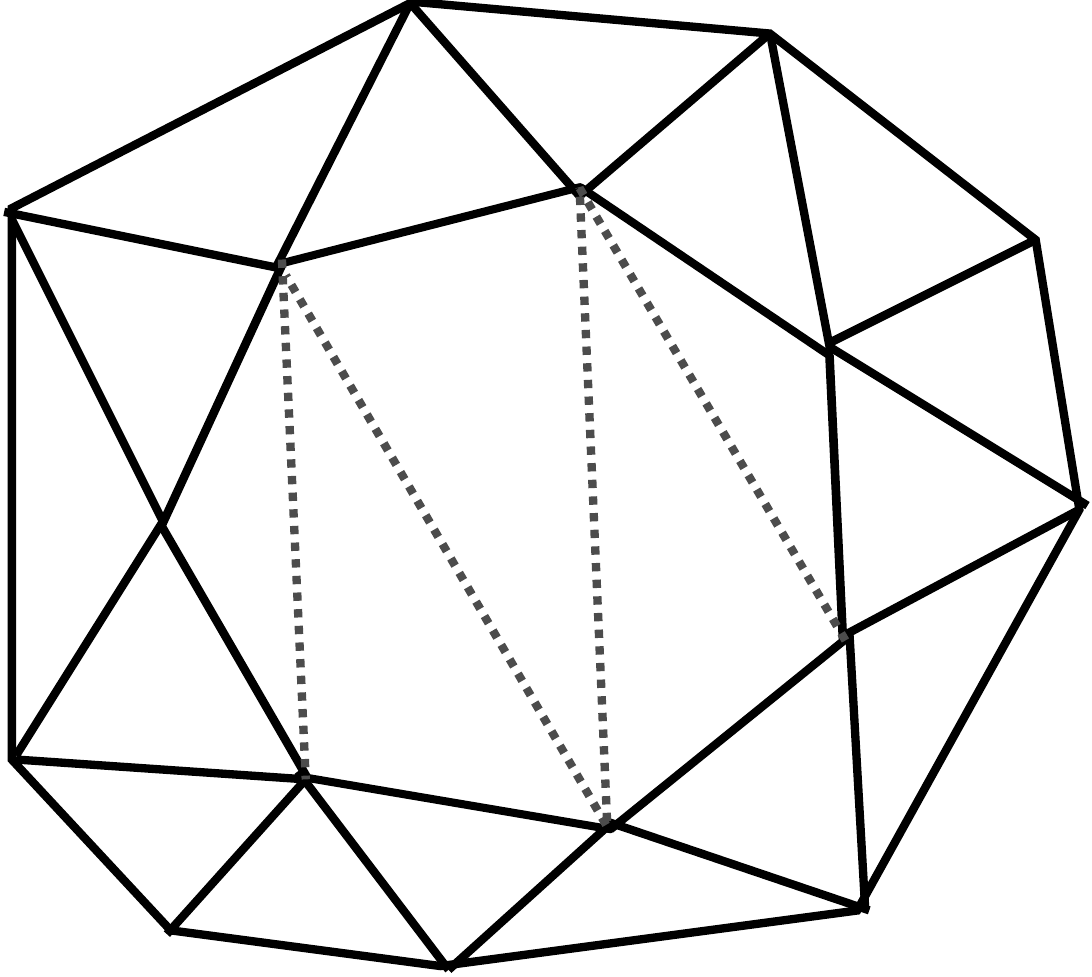}
\caption{\label{disconnex}Topology repair after node failure.}
\end{center}
\end{figure}

We make the assumption that a node leaving the network is equivalent to a failure of this node. A failure on a node $n$ is detected by its neighbors when they do not receive a ping message of $n$ within a given timeout. If a failure is detected, then the topology needs to be repaired in order to maintain a triangular mesh. When the failure of a node is detected by its neighbors, the node having the lowest identifier launches a repair agent. This is possible because each neighbor of the failing node is aware of all other neighbors. This agent creates connexions between neighbors of the node that has left, repairing the hole in the topology as shown in figure \ref{disconnex}.

There are two kinds of holes in the mesh: some are repairable whereas others are not. If the missing node has at least one neighboor with a degree of four, then the hole in the mesh cannot be fixed because the previously described procedure may lead to triangles with nodes having a degree of three, breaking the invariant property described in section \ref{invariant}. In this case, a random walker is sent through the network to find a node that can replace the failing node. The replacing node must leave a repairable hole in the mesh, which means that all of its neighbors must have a minimal degree of five.

Even if all neighbors of the failing node still have a degree of four, the repair agent may break the topological invariant. If the repair agent does not have a solution to repair the hole, it breaks all the connections previously created and another node in the network is moved to fill the gap.

\subsubsection{Topology flattening\label{flattening}}
Local deformation of the topology reduces the injectivity radius of the spiral walk. Node connections or node failures can lead to a locally curved topology. A flat topology allows replicas to self-organize in a near-optimal solution according to their dispersion strategy. Therefore the topology needs to be dynamically and locally flattened. This flattening process is performed each time a node joins the network or after a topology repairement.

\begin{figure}[htbp]
\begin{center}
\includegraphics[scale=0.4]{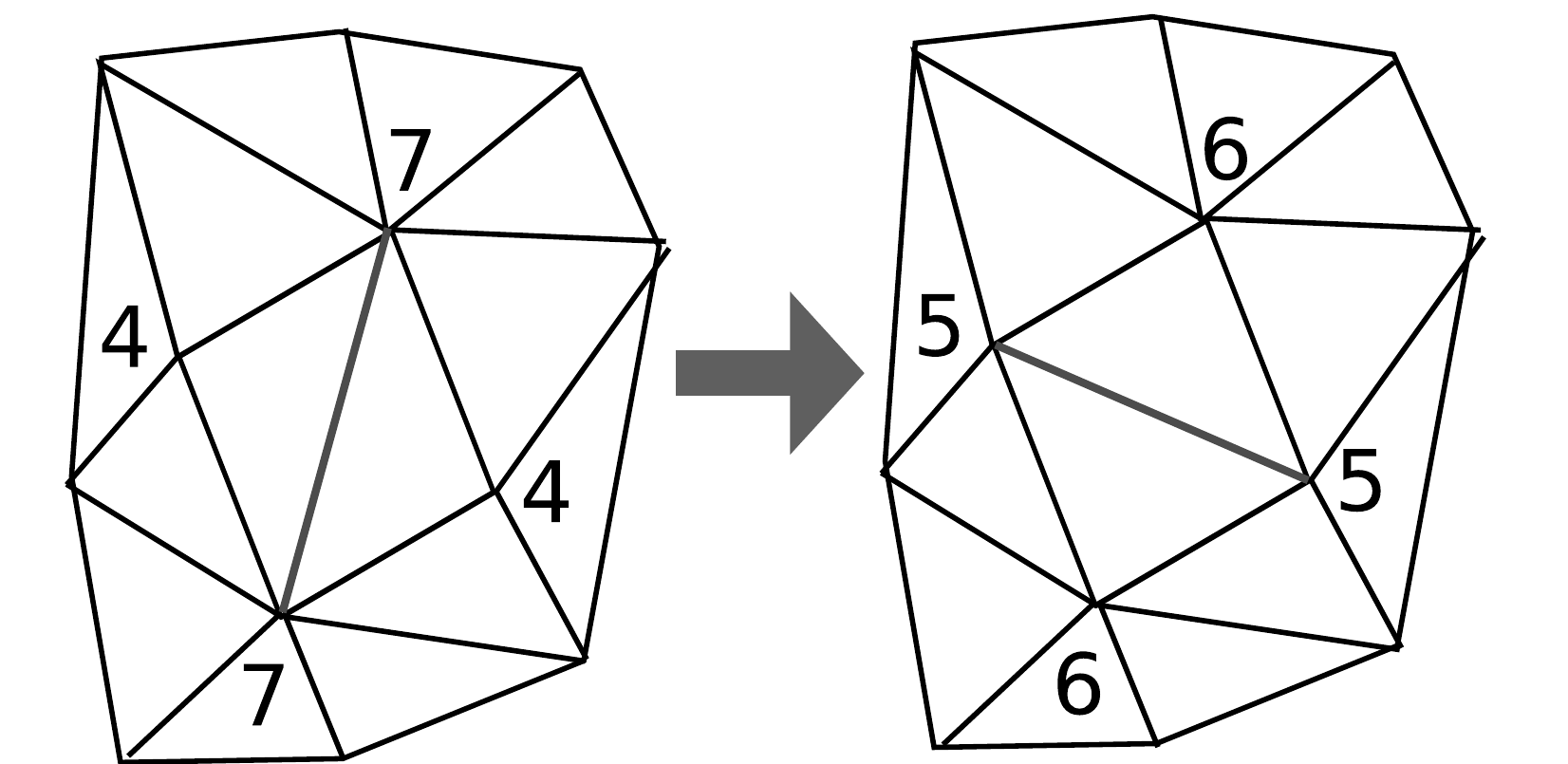}
\caption{\label{optimiz}Topology flattening.}
\end{center}
\end{figure}

For a given node, if it has a degree greater than six, then the neighbor with the highest degree is selected. If the sum of their degree is greater than the sum of the degree of their two common neighbors, then the two nodes break their connection and their two common neighbors connect together, as shown in figure \ref{optimiz}. This is performed only if the creation of the link between the two common neighbors does not break the topological invariant: the two common neighbors must have only two common neighbors.

When a node joins the network, it has a degree of four. This flattening process increases the degree of new nodes, and reduces the overall number of nodes having a degree of four. Flattening the network topology makes also the network repairment easier and simplifies the scanning of node at defined distance as described in the next section.

\subsection{Spiral Walk\label{sw}}
A biased random walk has been introduced by \cite{Huang2006}. This walk is designed for sensor networks which are a particular case of unstructured P2P networks. The bias introduced in this random walk favors the visit of nodes near the source before visiting those farther away. However, this kind of walk cannot guarantee that all nodes within a given radius will be visited.

In a sensor network, peers can connect to each other if they are close enough. In an unstructured P2P network built over internet topology, peers can connect to any other peer wherever it is, which makes it possible to control the virtual network topology. Using the algorithm proposed by \cite{Huang2006} on a regular triangular mesh (i.e. all nodes have a degree of six), one can notice that this algorithm performs a compact exploration. This kind of walk can perform very well when nodes have common neighbors.

We detail a spiral walk (SW) algorithm that performs a compact and exhaustive exploration of the neighborhood of a node on a triangular mesh, fullfilling the property described in section \ref{invariant}. Given a graph $G$ and a source node $n \in G$, SW visits all nodes in the neighborhood of $n$ within a given radius. This radius refers to the minimum number of hops of the shortest path between $n$ and other nodes.

The number of generated messages is linearly proportional to the number of visited nodes; unlike strategies such as flooding that generates a number of messages exponentially increasing with the radius of the exploration.

\subsubsection{Spiral walk principle}
Spiral walk visits all nodes within a given distance of the node that initiates the walk. Spiral walk can be viewed both as a local space filling curve and as a kind of\textit{ k-random walk}. It is a breadth-first search (BFS) exploration: this property ensures that the exploration within a given radius is performed exhaustively in a bounded number of hops.

This algorithm is designed for a triangular mesh on a closed 2-manifold with the following property: two neighboring nodes in the network have exactly two common neighbors, as described in section \ref{invariant}. This property allows the entire neighborhood of a node to be sequentially visited.

\begin{figure}[htbp]
\begin{center}
\includegraphics[scale=0.5]{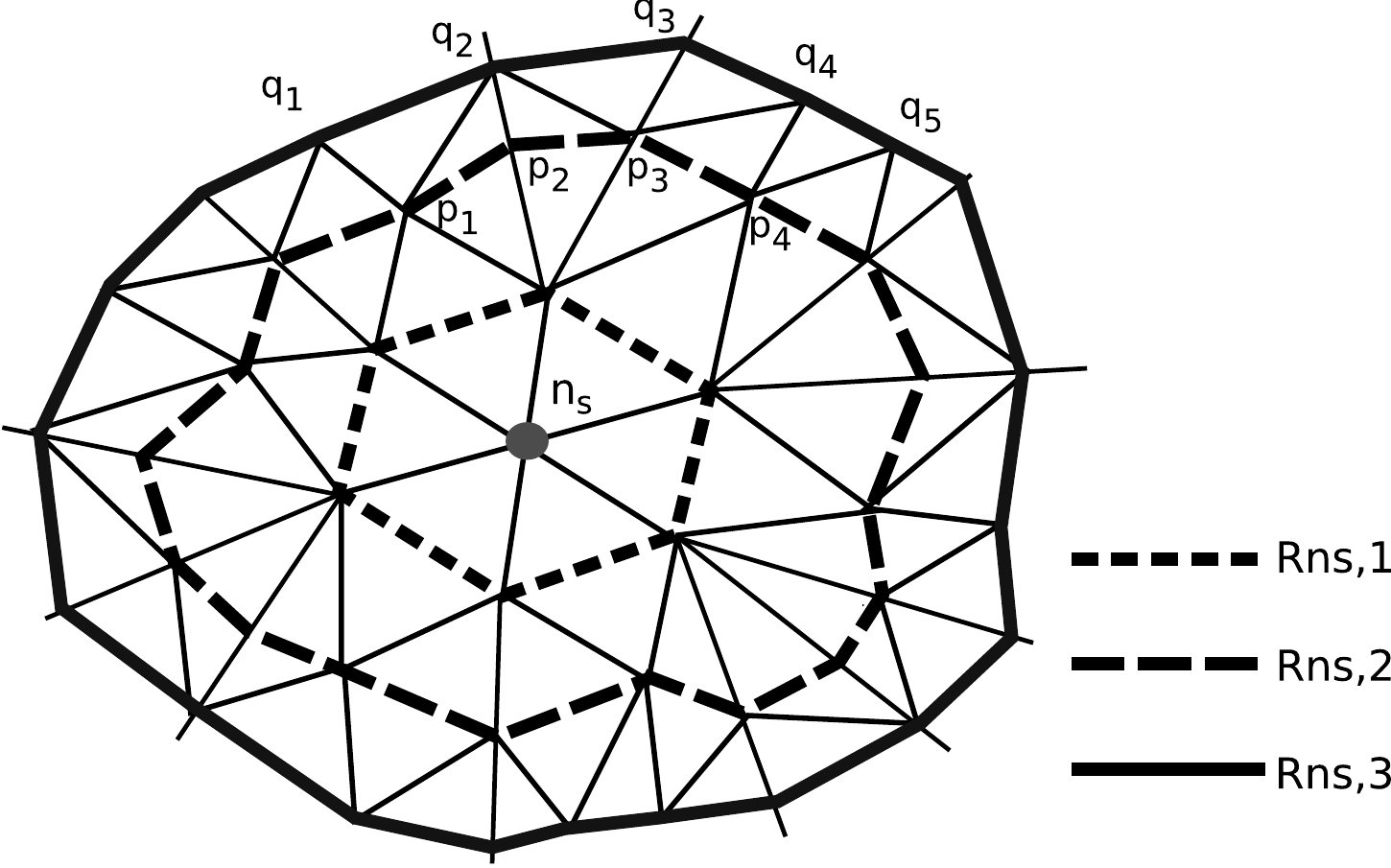}
\caption{Three nested rings of a spiral walk. $p_i$ is the sequence of nodes in $R_{n_s,2}$ used and $q_i$ is the sequence of visited nodes while exploring $R_{n_s,3}$.}
\end{center}
\end{figure}

We assume that the network topology is stable and that no node leaves or enters the network during the walk.

The main idea is to build rings starting from the node $n_s$ launching the walk. Each nested ring is built in reference to the previous one. Using the notations defined in section \ref{replicat}, $R_{n_s,i}$ is the $i^{th}$ ring of the spiral walk starting from $n_s$. $R_{n_s,0}$ is the first ring containing only the node $n_s$ initiating the walk. $R_{n_s,max}$ is the last ring of the walk, $max$ being the radius of the walk.

Having the following property: $\forall n \in R_{n_s,i-1}, |R_{n,1}\cap R_{n_s,i-1}|=2$, nodes in $R_{n_s,i-1}$ can be sequentially visited, since two neighbors $\in R_{n_s,i-1}$ have one common neighbor in $R_{n_s,i-2}$ and one common neighbor in $R_{n_s,i}$. The way $R_{n_s,i+1}$ is built, knowing $R_{n_s,i-1}$ and visiting $R_{n_s,i}$ is described in algorithm \ref{spiral}.

\begin{algorithm}[htbp]
\caption{Building next ring. \label{spiral}}
\begin{algorithmic}
\STATE $p\leftarrow n\ :\ n\in R_{n_s,i-1}, |R_{n,1}\cap R_{n_s,i}|>1$
\STATE $q\leftarrow n\ :\ n\in R_{n_s,i}\cap R_{p,1}, |R_{n,1}\cap R_{n_s,i-1}|>1 $
\STATE $P\leftarrow R_{n_s,i-1}$
\STATE $R_{n_s,i+1}\leftarrow\emptyset$
\WHILE{$P\neq\emptyset$}
\STATE $R_{n_s,i+1}\leftarrow R_{n_s,i+1}\cup \{n : n\in R_{q,1}\setminus R_{n_s,i-1}\cup R_{n_s,i}\}$
\STATE $q\leftarrow n\ :\ n\in R_{n_s,i}\cup R_{q,1}\cup R_{p,1}$
\IF{$|R_{q,1}\cap R_{n_s,i-1}|>1$}
\REPEAT
\STATE $P\leftarrow P\setminus\{p\}$
\STATE $p\leftarrow n\ :\ n\in  R_{n_s,i-1}\cap R_{p,1}\cap R_{q,1}$
\UNTIL{$|R_{p,1}\cap R_{n_s,i}|>1$}
\ENDIF
\ENDWHILE
\end{algorithmic}
\end{algorithm}

Each time a node $\in R_{n_s,i}$ is visited, all its neighbors that are not in $R_{n_s,i-1}$ or $R_{n_s,i}$ are recorded in $R_{n_s,i+1}$. When $R_{n_s,i}$ has been exhaustively visited, all $R_{n_s,i+1}$ are known. Spiral walk ends when the walker reaches a given TTL (Time To Live) given in time or hops units, or when the walker reaches a given radius.

\subsubsection{Walker's return\label{return}}
After having completed the exploration, the spiral walker returns to the initial node by a new way. During its exploration, it records in a hashtable the hop distance to the source of each node it visits. Using this information it can go back to the source using this gradient of distance. The length of the return path is equal to the radius of the walk.

Another interesting feature is that this property can be used to send back walker each time interesting information is found, so that source node can receive answers before the spiral walker finishes its exploration. As back walkers do not necessarily take the same path, this can give information gradualy, spreading network traffic over the whole neighborhood of the source node.

This generates a low number of messages everytime information is back propagated. Let $k$ be the number of answers and $N$ the length of the walk to find those $k$ answers. Because of the short cut, the number of message using spiral walker will be $O(k\sqrt{N})$. On the other hand, using random walk, the number of messages will be $O(k.N)$.

\subsubsection{Locally curved space\label{eye}}
\begin{figure}[htbp]
\begin{center}
\includegraphics[scale=0.3]{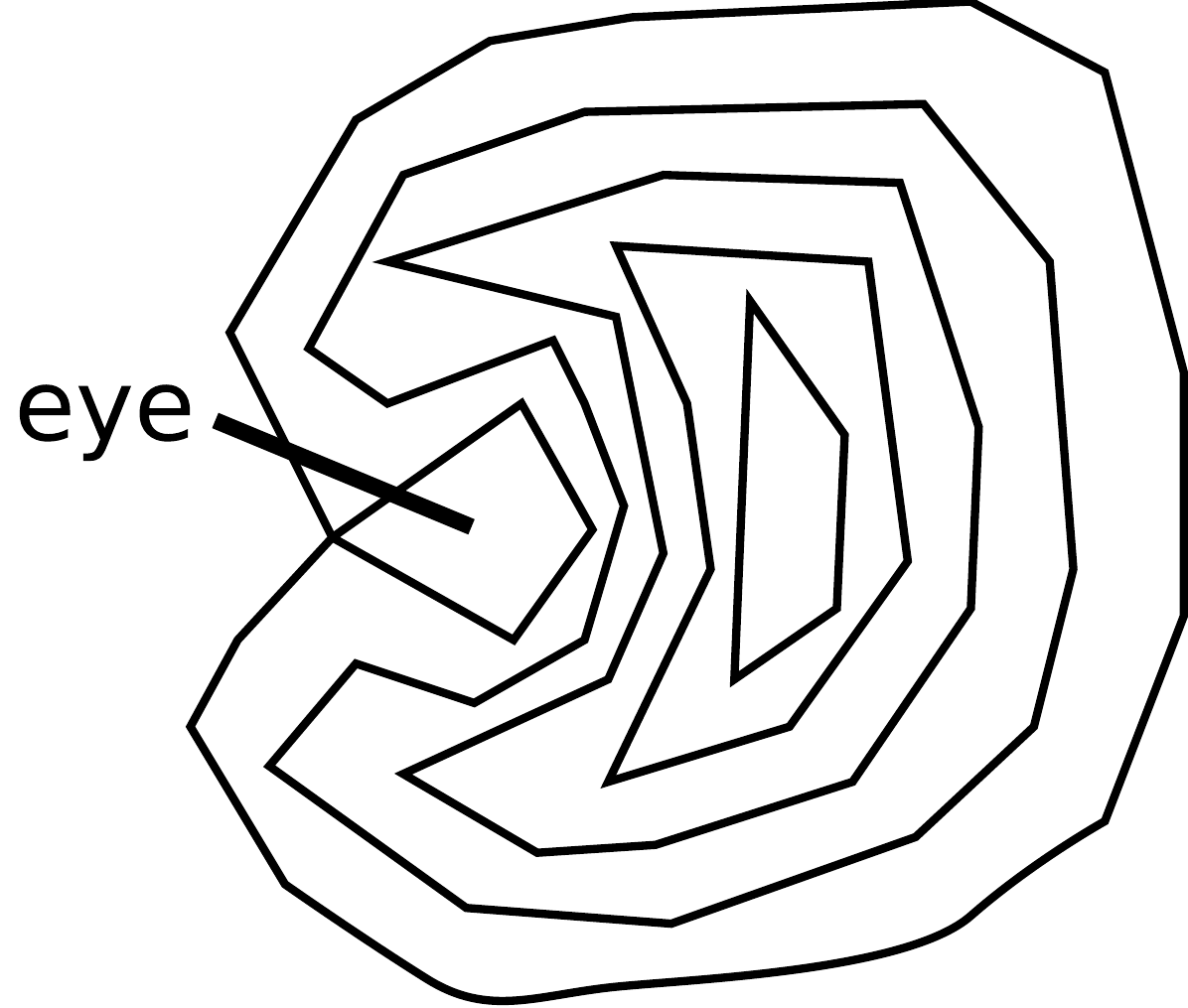}
\caption{Eyes appearing during the walk. \label{eyeFig}}
\end{center}
\end{figure}

The rules defined for node connection and the process involved in the topology repairement are very simple and the cost to maintain this topology is very low. However, these updating rules can lead to a curved space that reduces the injectivity radius of the walk: some nodes may be both distant in the ring and neighboors in the topology. This creates shortcuts for the building of the next ring and may lead to unexplored area, we call these phenomenons \textit{eyes}, as shown in the figure \ref{eyeFig}.

Although we present a way to reduce the curvature of the topology, \textit{spiral walk} may still visit the same node twice, especially if the radius of the walk is large. Because all nodes in $R_{n_s,i}$ are visited sequentially and all nodes in $R_{n_s,i-1}$ are known when $R_{n_s,i}$ is visited, we can detect if a node in $R_{n_s,i}$ has more than two neighbors in $R_{n_s,i}$. If it occurs, then this means that the space is locally curved. This problem is fixed by spawning spiral walkers accross these eyes.

Each spawned spiral walker inherits the memory of the original one and can go back to the source node that has initiated the walk using the gradient of distance as described in section \ref{return}. The only difference between the two spiral walkers is that the former will build larger and larger rings, whereas the latter will build smaller and smaller ones. This allows us to deal with extreme configurations of the topology where some eyes may recursively contain other eyes.

When these configurations occur, our algorithm can be viewed as a kind of \textit{k-walker}, with $k-1$ eyes. The advantage of this approach is that, thanks to the properties of the network topology and the gathering of information about visited nodes, all spawned walkers will not visit the same nodes, whereas in traditional \textit{k-random walk} this may happen.

It creates a parallel search, each spawned walkers having its own domain to explore, but it alters replicas placement and leads to a worse organization than on a flat topology.

\subsubsection{Dealing with very dynamic environment}
Previously, we made the assumption that the network is stable during a spiral walk. Here, we show how change of the network topology during the walk can be dealt with. The solution provides an approximative walk and the exhaustivity of the walk is lost.

The first solution uses the previously described \textit{spiral walk} algorithm. Each time an interesting piece of information is found, a walker is sent back. When the walker detects an anomaly in the building of its rings, it stops. This solution provides a partial answer if a node in $R_{i-1}$, $R_i$ or $R_{i+1}$ disconnects while the walker is in $R_i$. Otherwise, the algorithm still works as it uses a gradient mecanism very resilient to node departure to go back to the source node that initiated the spiral walk.

The second solution uses the algorithm proposed by \cite{Huang2006}. This walk can be performed in very dynamic environments. However, because of its random behavior in non dedicated network topologies, it does not ensure the exhaustivity of the walk and reduces the efficiency of the replication algorithm.


\section{Search protocol}
\subsection{Introduction}
The cost of random walk and variants for the routing of queries is generally of $O(log(N))$ for a network with $N$ nodes. This seems to be very high when compared to most standard routing P2P mechanisms, but the difference is mainly related to the heuristics used. Since no assumption is made in our case for the kind of query performed, the routing cost is balanced by the fact that ranged or complex queries can be performed along the walker’s path.

Using the super-peer paradigm, as suggested by \cite{Yang.ICDE.2003}, can speed up the search by a factor proportional to the average number of sub-peers per super-peers. For instance, if one percent of nodes in the network have $100$ times more resources than the average of nodes, the search space can be reduced by a factor of $100$.
In the following sections, we describe the layout of our system and introduce the rules for the creation of super-peers.

%

\subsection{Search layout}
A super-peer is a node in the network that has above-average resources. A super-peer is connected to other super-peers, allowing to efficiently forward queries in the whole network, assuming the super-peer network is strongly connected. Moreover, a super-peer manages many sub-peers, the number of managed sub-peers being proportional to the super-peer resources. In most existing unstructured P2P networks, sub-peers are called clients because they are only connected to super-peers. We use the term sub-peers because all nodes in the networks are not connected to only super-peers, but also to sub-peers.

\begin{figure}[htbp]
\begin{center}
\includegraphics[scale=0.5]{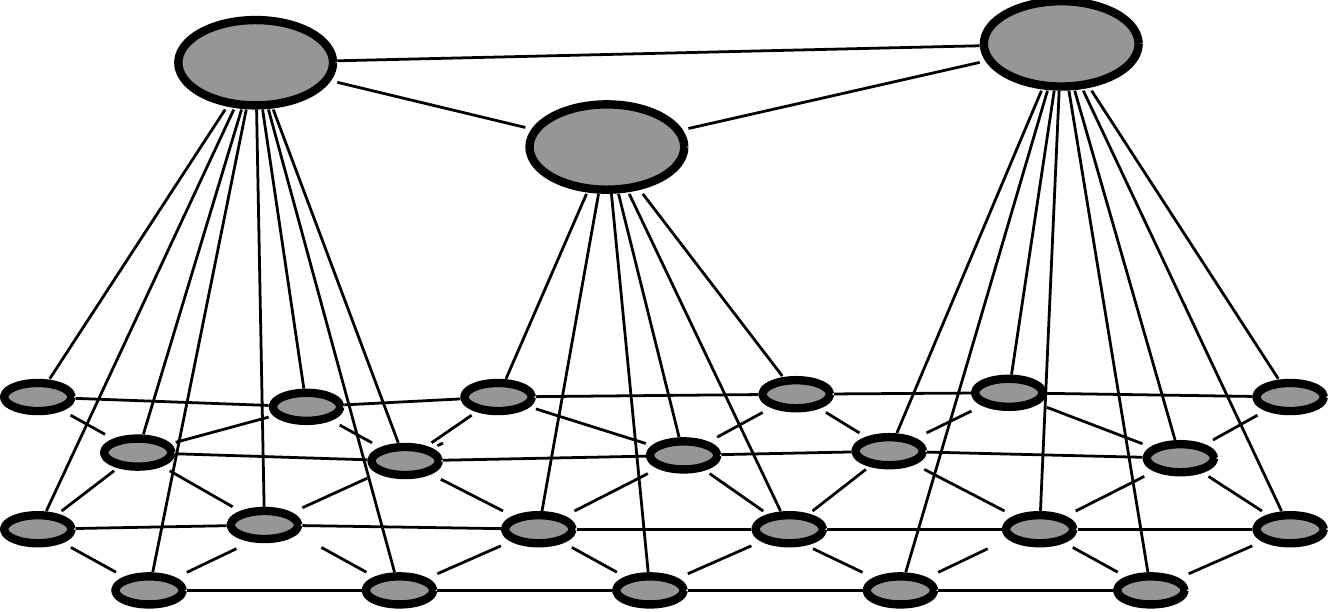}
\caption{System design\label{topo} : the triangular mesh allows compact exploration at low cost while high connectivity nodes are used to increase search efficiency.}
\end{center}
\end{figure}

Each super-peer is responsible for at least one other peer (including itself) and each peer has one super-peer, as shown in figure \ref{topo}.

\subsection{Query routing}
In order to speed up query routing, each node indexes the data locally stored and sends a copy of the indexed information to its related super-peer.

A query initiated by a peer is forwarded to its super-peer, with a Time To Live (TTL) strictly positive. The super-peer first handles the query for each of its other sub-peers. Then TTL is reduced by one and the query is forwarded to a random neighbor in the super-peer layer. Each time a super-peer receives a query, the query is handled for its sub-peers and forwarded to another random neighbor in the super-peer layer until its TTL reaches $0$.

\subsection{Super peer management rules}
Super-peers are the entry points for nodes wishing to join the network. It can be difficult for a node to contact a super-peer of the network, that is why super-peers never refuse new clients. To avoid being overloaded, super-peers regulate their number of sub-peers themselves.

\subsubsection{Super-peer creation\label{supercreate}}
If a super-peer $n_1$ has to manage too many sub-peers, one of its sub-peers $n_2$ is promoted super-peer to rebalance a too heavy load. The sub-peer selected is the one having the highest capabilities among all sub-peers of $n_1$. The new super-peer $n_2$ contacts its nearest neighbors in the triangular mesh and becomes their new super-peer. The number of nodes contacted is proportional to the resources of $n_2$. Nodes are contacted by launching a spiral walker in the triangular mesh.

This scheme ensures that all sub-peers of $n_2$ are close in the triangular mesh. Because replicas of a given piece of data repulse each other in the triangular mesh, a super-peer will unlikely have twice the same piece of data from different sub-peers.

In order to efficiently speed up information retrieval and to keep the super-peer network resilient to fragmentation, the new super-peer $n_2$ has to aquire new neighbors in the super-peer network. This is achieved by sending a random walker in the super-peer network through $n_1$. Super-nodes visited by the random walker contact $n_2$ and become its new neighbor in the super-peer network. Only super-peers having a low degree in the super-peer network try to connect to $n_2$, as the failure of high-degree super-peers may lead to a fragmented network.

\subsubsection{Super-peer failure}
We make the assumption that a super-peer never notify its departure from the network. When a super-peer loses a neighbor in the super-peer network and its degree becomes low, it can acquire new super-peer neighbors, using the strategy described previously for super-peer creation.

However, the absence of notification at the sub-peer level raises problems for the sub-peers of the failing super-peer because they cannot send their queries accross the super-peer network. Since they do not have a super-peer, these nodes launch their queries accross the triangular mesh. Those queries are forwarded randomly in this part of the network until a node having a super-peer is encountered. The query is then forwarded at random in the super-peer network.

\begin{algorithm}[htbp]
\caption{Regulation from super node $sn_1$\label{regulation2}}
\begin{algorithmic}
\WHILE{true}
\STATE $sn_2 \leftarrow sn_1.getRandomSuperNeighbor()$
\IF{$\frac{sn_1.numSubPeers()+sn_2.numSubPeers()}{sn_1.capabilities()} < 1$}
\FORALL{$n \in sn_2.getSubPeers()$}
\STATE $sn_1.addSubPeer(n)$
\STATE $n.setSuperNode(sn_1)$
\ENDFOR
\STATE Downgrade $sn_2$ as normal peer.
\IF{$sn_1.hasFewSuperNeighbors()$}
\STATE Find new super neighbors for $sn_1$.
\ENDIF 
\ENDIF
\STATE Wait $x$ sec.
\ENDWHILE
\end{algorithmic}
\end{algorithm}

Because the sub layer is strongly connected, some sub-peers of the failing super-peer have neighbors in the topology with a different super-peer, as described in \ref{regul}. We call such sub-peers \textit{borderline-peers}. When they lose their super-peer, all \textit{borderline-peers} create a link to the super-peer of one of their neighbors. All their neighbors which have lost their super peer and are not borderline peers become \textit{borderline-peers}, as they now have a neighbor in the triangular mesh with a super-peer.

\subsubsection{Load balancing\label{regul}}
When there are more departures than arrivals in the network, the overall number of node decreases. Super-peers have on average a higher uptime than other peers. To keep the number of super-peers proportional to the number of peers on average, there is a need to downgrade super-peers. This is achieved by the algorithm \ref{regulation2}: neighboring super-peers periodically send a message containing their load measure to each other. If one of them can handle the whole load, it recovers all sub-peers of its neighbor and the other is downgraded as normal-peer.

As shown in figure \ref{topo}, all nodes have a set of neighbors in the sub layer of the network. Two neighbors in the sub layer graph can have a different super-peer. In this configuration, the two neighbors periodically compare the load measure of their super-peer. They can become the sub-peer of the super-peer of their neighbor to balance the load between super-peers, as described in algorithm \ref{regulation1}.

\begin{algorithm}[htbp]
\caption{Bottom-up regulation from node $n$\label{regulation1}}
\begin{algorithmic}
\WHILE{true}
\STATE $sn_1 \leftarrow n.getSuperNode()$
\STATE $sn_2 \leftarrow n.getRandomNeighbor().getSuperNode()$
\STATE $load_1 \leftarrow \frac{sn_1.numSubPeers()-1}{sn_1.capabilities()}$
\STATE $load_2 \leftarrow \frac{sn_2.numSubPeers()+1}{sn_2.capabilities()}$
\IF{$load_1 > load_2$}
\STATE $sn_1.removeSubPeer(n)$
\STATE $n.setSuperNode(sn_2)$
\STATE $sn_2.addSubPeer(n)$
\ENDIF
\STATE Wait $x$ sec.
\ENDWHILE
\end{algorithmic}
\end{algorithm}


\section{Experiments\label{experiments}}
Experiments have been performed on a dedicated simulator we have developped in Java. At initialization, the network is made of 12 nodes and the topology is an icosahedron, with all nodes having a degree of five. The average node degree gets closer to six as more nodes are added to the network.

\begin{figure}[htbp]
\begin{center}
\includegraphics[scale=0.45]{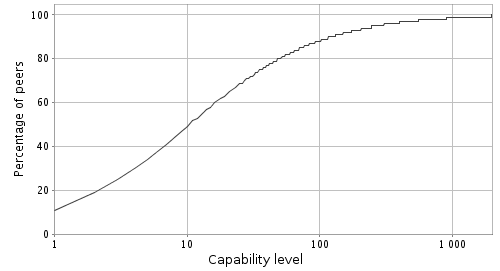}
\caption{Simulated capability repartition among peers in the network.\label{capabilities}}
\end{center}
\end{figure}

In order to simulate heterogenity in peers, nodes are given capabilities following the distribution in figure \ref{capabilities}. These capabilities reflect the number of connexions or load of work a node can handle. We believe this distribution is close to real distribution, as shown by the measurements in \cite{saroiu03measuring,1217970}. More precisely, this capability is used to compute the amount of data a node can host, or the number of sub-peers it can manage when promoted super-peer.

\subsection{Replication\label{replicationE}}
To evaluate the influence of the regularity of the Adapnet mesh, we compare the results with the same scheme performed on an optimal mesh. By recursively subdividing an icosaedron, we can generate a triangular geode. This can be viewed as an optimal closed triangular mesh mapped on a sphere. In this section, we refer as the experimental topology the network obtained using the protocols described in section \ref{topology} for node connection or departure. The optimal topology, starting with an icosaedron, is a triangular geode. In order to perform a comparison, the geode we have simulated contains $40962$ nodes, so does the simulated experimental topology.

\begin{figure}[htbp]
\begin{center}
\includegraphics[scale=0.45]{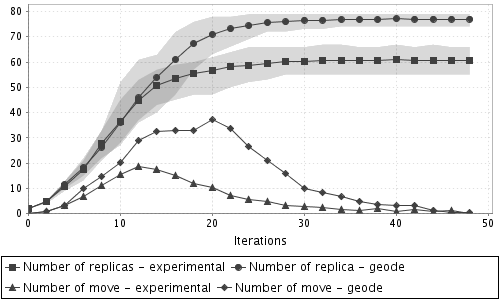}
\caption{Evolution of the quantity of replicas for one piece of data. $r=25$, $maxScore=250$ and $t=16$.\label{repli}}
\end{center}
\end{figure}

The graph in figure \ref{repli} shows the evolution of the number of replicas for one piece of data on the experimental topology and on the geode. Both topologies are static, no node joins or leaves the network. We assume in this experiment that there is always some available space in the cache of nodes and that caches are never purged.

Initially, a piece of data is put on a random node. Then during each iteration, each node in the network having a replica of the data scans its neighborhood and decides if it must create another replica, move the one it is hosting on another node, delete it, or do nothing , as described in algorithm \ref{replication}.

We can see on figure \ref{repli} that, given a stable topology, the replication algorithm converges. The number of replicas created or moved drops to zero. More replicas are created on the geode: this is because the geode has a higher diameter than the topology in practice. Irregularities on the topology make the algorithm converge faster, we believe it is because it reduces the number of sub-optimal configurations.

\begin{figure}[htbp]
\begin{center}
\includegraphics[scale=0.45]{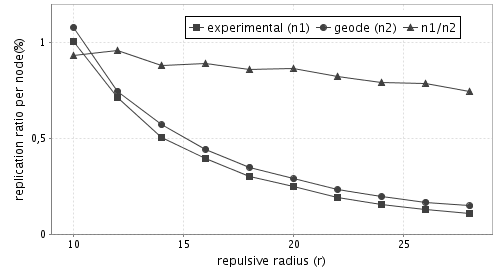}
\caption{Quantity of replicas according to the repulsive radius $r$ for one piece of data.  $maxScore=10\times r$ and $t=4$.\label{repul}}
\end{center}
\end{figure}

Figure \ref{repul} shows the impact of the repulsive radius on the global amount of replicas. It confirms that a perfect mesh has a higher diameter than an irregular one, given the same amount of nodes. As a consequence, less replicas are requiered to populate the whole network. As the repulsive radius $r$ increases, the geode topology proportionally includes more replicas than the experimental topology.

Irregularities into the experimental topology lowers the amount of required replicas to populate the network for a given repulsing radius.

\begin{figure}[htbp]
\begin{center}
\includegraphics[scale=0.47]{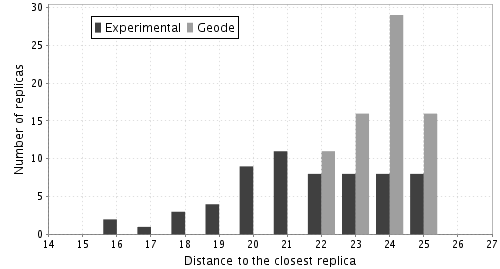}
\caption{Source repartitions after replication with $r=25$, $maxScore=250$ and $t=16$.\label{reparti}}
\end{center}
\end{figure}

Figure \ref{reparti} shows how well the replicas are spread accross the network. We measure the distance between the node hosting a replica and the closest node in the network hosting the same replica. This measurement has been collected just after a replication procedure has been performed.

Replicas are better distributed on the geode topology. Nevertheless, the experimental topology exhibits a density mostly confined into a small bounded interval ($[20,25]$). This shows the effectiveness of the replication procedure applied on our dedicated topology.

\subsection{Adaptation to the number of nodes\label{adaptation}}
We simulate failure or entering of new nodes to artificially increase or decrease the total number of nodes and observe the evolution of the number of replicas for one piece of data.

\begin{figure}[htbp]
\begin{center}
\includegraphics[scale=0.47]{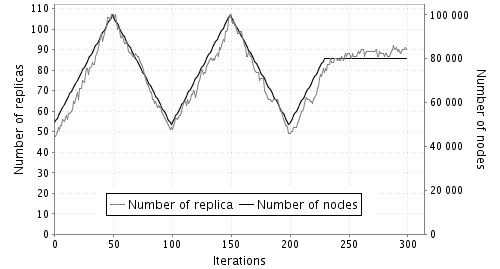}
\caption{Evolution of the number of replicas according to the network's size. $MaxScore = 10\times r$.\label{var1}}
\end{center}
\end{figure}

\begin{figure}[htbp]
\begin{center}
\includegraphics[scale=0.47]{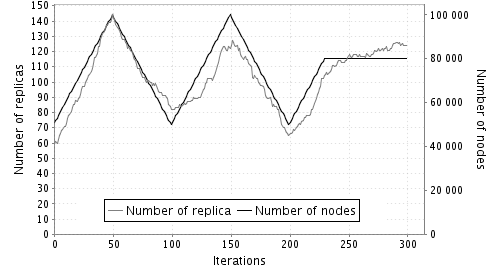}
\caption{Evolution of the number of replicas according to the network's size. $MaxScore = 20\times r$.\label{var2}}
\end{center}
\end{figure}

Figures \ref{var1} and \ref{var2} show the evolution of the number of replicas (bright line) according to the entering or departure of nodes in the network (dark line). Setting used are $r = 30$ and $t=16$. Again, we suppose that there is always enough space in the cache of nodes and that caches are never purged.

Although there are small oscillations, we can see that the overall quantity of replicas remains proportional to the number of nodes in the network. As awaited, $maxScore$ controls the variation of reactivity of the replication algorithm described in section \ref{replischeme}.

\subsection{Replication of rare information\label{repli_full}}
In this experiment, we have generated  a $10\ 000$ nodes network, storing data on nodes according to a distribution close to the one measure by \cite{1217970}: few data is highly replicated while the majority of data has none or few replicas. On the $100\ 000$ items spread across the network, half of them are not replicated and two third have at most one replica. All items have the same size. Each cache is sized according to the number of item allocated for a node. Setting used are $r = 30$ and $t=4$.

\begin{figure}[htbp]
\begin{center}
\includegraphics[scale=0.47]{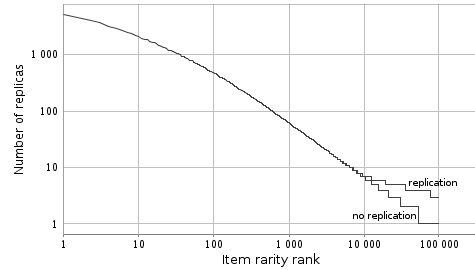}
\caption{Evolution of the number of replica for each item in the network. Items are ranked by increasing rarity. Only rare data is replicated.\label{dataRepartition}}
\end{center}
\end{figure}

The figure \ref{dataRepartition} confirms that only rare data is replicated. For highly replicated items ($10$ replicas or more), the amount of replicas remains the same.

\subsection{Answering speed}
Since super-peers hold indexed information of their sub-peers, we measure the answering speed by counting the number of super-peers visited by a random walker in the super-peer network before the query is answered. We generate a $100\ 000$ nodes network with the experimental topology and spread one item and its replicas accross the network.

The fact that it is quite unlikely that a super-peer hosts several copies of the same piece of data indexed by distinct sub-peers, slighly increases the answering speed (section \ref{supercreate}). In this experiment, the number of sub-peers a super-peer can manage is $10$ times less than its capability. Using distribution of figure \ref{capabilities}, the simulated network contains on average $1.2 \%$ of super-peers.

\begin{figure}[htbp]
\begin{center}
\includegraphics[scale=0.47]{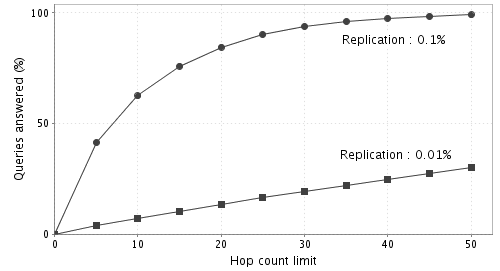}
\caption{Query answer speed.\label{recup}}
\end{center}
\end{figure}

The figure \ref{recup} shows the number of queries fullfilled according to the maximum number of nodes that can be visited by the walker (hop count limit). For a $100\ 000$ nodes network containing on average $1.2 \%$ of super-peers, the search space is reduced to $1200$ nodes. The use of super-peers greatly increases answering speed, even if low replicated data are still difficult to locate.


\section{Conclusion and future work}
We introduced a data P2P architecture that uses storage space offered by users to increase data availability through wise replication. The replication scheme relies on exhaustive exploration of the neighborhood of each node, adapting dynamically the number of data replica. The agent-based regulation mechanisms presented give the system some homeostasis properties: the global number of replicas remains stable and wisely dispatched whithout any supervised nor global awarness or the network state when a node enters or leaves the network. The spiral walk algorithm is supported by a near-regular triangular closed mesh : this topology is maintained by a rule-based system that keeps repairing the mesh locally.

The search cost introduced by the fact that the system makes no assumption on the type of query used, is lowered by a dedicated system layout based on the use of super-peers.

We have presented an exploration scheme that generates a number of messages lineary proportional to the number of visited nodes. However, when the topology is well suited for replication, it generates one walker, whereas it become parallelised only when the topology is not suited for replication. We are currently looking toward the possibility to launche many walkers in parallel that will explore their own area of the neighborhood of any anode. This parallelisation could speed up neighborhood exploration, reducing the possibility of node departure while performing the exploration, thus increasing system robustness.

Another interesting feature that could be added is the use of erasure codes \cite{Wicker94} for data replication in the cache offered by users.


\bibliographystyle{plain}
\bibliography{adapnet}
\end{document}